\documentclass [12pt] {article}
\usepackage{latexsym}
\usepackage{latexsym}
\newtheorem{theorem}{Theorem}
\newtheorem{proposition}{Proposition}

\newtheorem{lemma}{Lemma}

\title{On Stability Property of Probability Laws with Respect to Small
Violations of Algorithmic Randomness}
\author{Vladimir V. V'yugin\thanks{
This paper is an extended version of the talk at the
Eighth International Conference on Computability, Complexity and
Randomness (CCR 2013). September 23-–27, 2013, Moscow, Russia;
see also the conference paper V'yugin~\cite{Vyu2011}. This work is
partially supported by grant RFBR 13-01-12458}.
\\
{\small Institute for Information Transmission Problems}
\\
{\small and}
\\
{\small National Research University Higher School of Economics}
\\
{\small Bol'shoi Karetnyi per. 19, Moscow GSP--4, 127994, Russia}\\
{\small e-mail vyugin@iitp.ru}
}

\date{}

\begin{document}
\maketitle

\begin{abstract}
We study a stability property of probability laws with respect
to small violations of algorithmic randomness. A sufficient
condition of stability is presented in terms of Schnorr tests
of algorithmic randomness. Most probability laws, like the
strong law of large numbers, the law of iterated logarithm, and
even Birkhoff's pointwise ergodic theorem for ergodic
transformations, are stable in this sense.

Nevertheless, the phenomenon of instability occurs in ergodic theory.
Firstly, the stability property of the Birkhoff's ergodic theorem
is non-uniform. Moreover, a computable non-ergodic measure preserving
transformation can be constructed such that ergodic theorem is non-stable.
We also show that any universal data compression scheme is also non-stable
with respect to the class of all computable ergodic measures.
\end{abstract}

\section{Introduction}

In this paper we study stability property of probability laws
with respect to small violations of randomness. By a
probability law we mean a property $\Phi(\omega)$ of infinite
binary sequences $\omega$ which holds almost surely. We define a
notion of stability of a probability law in terms of algorithmic theory of
randomness. Within the framework of this theory the probability
laws are formulated in ``a pointwise'' form.
It is well known that main laws of probability theory are valid
not only almost surely but for each individual Martin-L\"of random sequence.

Some standard notions of algorithmic randomness are given in Section~\ref{random-ppp1}.
We use the definition of a random sequence in the complexity terms.
An infinite binary sequence $\omega_1\omega_2\dots$ is
Martin-L\"of random with respect to uniform (or $1/2$-Bernoulli) measure
if and only if $Km(\omega^n)\ge n-O(1)$ as $n\to\infty$,
where $Km(\omega^n)$ is the monotonic Kolmogorov complexity of
a binary string $\omega^n=\omega_1\dots\omega_n$ and the constant $O(1)$
depends on $\omega$ but not on $n$.
\footnote
{The same property holds also if we replace monotonic complexity $Km(\omega^n)$
on the prefix complexity $KP(\omega^n)$. The difference is that an inequality
$Km(\omega^n)\le n+O(1)$ holds for monotonic complexity
but this is not true for prefix complexity.}

A probability law $\Phi(\omega)$ is called stable if an unbounded
computable function
$\sigma(n)$ exists such that $\Phi(\omega)$ is true for each infinite
sequence $\omega$ such that $Km(\omega^n)\ge n-\sigma(n)-O(1)$ as $n\to\infty$.
We assume that this function non-decreasing and refer to the function
$\sigma(n)$ as to a degree if stability.

A stability property under small violations of algorithmic randomness
of the main limit probability laws was discovered by Schnorr~\cite{Sch71} and
Vovk~\cite{Vov87}. They shown that the law of large numbers for the
uniform Bernoulli measure holds for a binary sequence $\omega_1\omega_2\dots$
if $Km(\omega^n)\ge n-\sigma(n)-O(1)$, where $\sigma(n)$ is an arbitrary computable
function such that $\sigma(n)=o(n)$ as $n\to\infty$, and
the law of iterated logarithm holds if $Km(\omega^n)\ge n-\sigma(n)-O(1)$,
where $\sigma(n)$ is an arbitrary computable function such that
$\sigma(n)=o(\log\log n)$.
\footnote
{
In what follows all logarithms are on the base 2.
}
V'yugin~\cite{Vyu97a} shown that the law of the length of longest head-run
in an individual random sequence is stable with degree of stability
$\sigma(n)=o(\log\log n)$.
It was shown in these papers that corresponding degrees of stability
are tight.

We present in Proposition~\ref{suff-1} a sufficient condition of stability
in terms of Schnorr tests of randomness. We mention that if a computable
rate of convergence almost surely exists then the corresponding probability
law holds for any Schnorr random sequence. In turn, the latter property
implies stability property of this law. Using this sufficient condition, we
prove that most probability laws, like the
strong law of large numbers and the law of iterated logarithm, are stable
under small violations of algorithmic randomness.
Theorem~\ref{stability-1d} shows that the Birkhof's ergodic theorem is
also stable in case where measure preserving transformation is
ergodic.

In Section~\ref{instab-1} we show that the phenomenon of
instability occurs in ergodic theory. First, there are no
universal stability bounds in ergodic theorems or ergodic
transformations. The Birkhof's ergodic theorem is non-stable
for some non-ergodic stationary measure preserving
transformation.

We note that there is some analogy
with the lack of universal convergence rate estimates in ergodic theory.
A lack of universal convergence bounds is typical for
asymptotic results of ergodic theory like Birkhoff's ergodic
theorem -- Krengel~\cite{Kre85}, Shannon--McMillan--Breiman
theorem and universal compressing schemes --
Ryabko~\cite{Ryb84}.

\section{Preliminaries}\label{random-ppp1}


Let $\Theta=\{0,1\}^*$ be a set of all finite binary sequences
(binary strings) and $\Omega=\{0,1\}^\infty$ be a set of all infinite
binary sequences. Let $l(\alpha)$ denotes the length of a
sequence $\alpha$ ($l(\alpha)=\infty$ for $\alpha\in\Omega$).

For any finite or infinite sequence $\omega=\omega_1\omega_2\dots$, we write
$\omega^n=\omega_1\omega_2\dots\omega_n$, where $n\le l(\omega)$.
Also, we write $\alpha\subseteq\beta$ if $\alpha=\beta^n$ for some $n$.
Two finite sequences $\alpha$ and $\beta$ are incomparable if
$\alpha\not\subseteq\beta$ and $\beta\not\subseteq\alpha$.
A set $A\subseteq\Theta$ is prefix-free if any two distinct sequences from $A$
are incomparable.

A complexity of a string $x\in\Theta^*$ is equal to the length of the shortest
binary codeword $p$ (i.e. $p\in\{0,1\}^*$) by which the string $x$
can be reconstructed:
$
K_\psi(x)=\min\{l(p):\psi(p)=x\}.
$
We suppose that $\min\emptyset=+\infty$.

By this definition the complexity of $x$ depends on a computable
(partial recursive) function
$\psi$ -- method of decoding. Kolmogorov proved that an
optimal decoding algorithm $\psi$ exists such that
$
K_\psi(x)\le K_{\psi'}(x)+O(1)
$
holds for any computable decoding function $\psi'$ and for all strings
$x$.
We fix some optimal decoding function $\psi$. The value
$K(x)=K_\psi(x)$ is called Kolmogorov complexity
of $x$.

If domains of decoding algorithms are prefix-free sets the same construction
gives us the definition of prefix complexity $KP(x)$.

Let $\cal R$ be a set of all real numbers, $\cal Q$ be a set of
all rational numbers.

A function $f:\Theta\to {\cal R}$ is called computable if there
exists an algorithm which given $x\in\Theta$ and a rational $\epsilon>0$
computes a rational approximation of a number $f(x)$ with
accuracy $\epsilon$.

For a general reference on algorithmic randomness, see
Li and Vit{\'a}nyi~\cite{LiV97}. We confine our attention to the
Cantor space $\Omega$ with the uniform Bernoulli measure $B_{1/2}$.
Hoyrup and Rojas~\cite{HoR2009} proved that any computable probability
space is isomorphic to the Cantor space in both the computable and
measure-theoretic senses. Therefore, there is no loss of generality
in restricting to this case.

The topology on the set $\Omega$ is generated by the binary intervals
$\Gamma_x=\{\omega\in\Omega:x\subset\omega\}$, where $x$
is a finite binary sequence.
An important example of computable measure is the uniform Bernoulli measure
$B_{1/2}$, where
$
B_{1/2}(\Gamma_x)=2^{-l(x)}
$
for any finite binary sequence $x$.

An open subset $U$ of $\Omega$ is called effectively open if it
can be represented as a union of a computable sequence of binary
intervals: $U=\cup_{i=1}^\infty\Gamma_{\alpha_i}$, where
$\alpha_i=f(i)$ is a computable function from $i$. A sequence
$U_n$, $n=1,2,\dots$, of effectively open sets is called
effectively enumerable if each open set $U_n$ can be represented as
$U_n=\cup_{i=1}^\infty\Gamma_{\alpha_{n,i}}$, where
$\alpha_{n,i}=f(n,i)$ is a computable function from $n$ and
$i$.

Martin-L\"of test of randomness with respect to a computable measure $P$
is an effectively enumerable sequence $U_n$, $n=1,2,\dots$, of effectively
open sets such that $P(U_n)\le 2^{-n}$ for all $n$. If the real numbers
$P(U_n)$ are uniformly computable then the test $U_n$, $n=1,2,\dots$, is called
Schnorr test of randomness.
\footnote{
Uniform computability of $P(U_n)$ means that there is an algorithm which given
$n$ and $\epsilon>0$ outputs a rational approximation of $P(U_n)$ up 
to $\epsilon$.
}

An infinite binary sequence $\omega$ passes a test $U_n$, , $n=1,2,\dots$,
if $\omega\not\in\cap U_n$. A sequence $\omega$ is Martin-L\"of random
with respect to the measure $P$ if it passes each Martin-L\"of
test of randomness. A notion of Schnorr random
sequence is defined analogously.

In what follows we mainly consider the notion of
randomness with respect to the uniform Bernoulli measure $B_{1/2}$.

An equivalent definition of randomness can be obtained using
Solovay tests of randomness.
A computable sequence $\{x_n:n=1,2,\dots\}$ of binary strings is called
Solovay test of randomness with respect to the uniform measure if the series
$\sum\limits_{n=1}^\infty 2^{-l(x_n)}$ converges.

An infinite sequence $\omega$ passes a Solovay test of randomness
$\{x_n:n=1,2,\dots\}$ if $x_n\not\subseteq\omega$ for almost all $n$.

We use an equivalence between
Martin-L\"of and Solovay tests of randomness.
\begin{proposition}\label{Sol-Mart-1}
An infinite sequence $\omega=\omega_1\omega_2\dots$ is Martin-L\"of random
if and only if it passes each Solovay test of randomness.
\end{proposition}
{\it Proof}. Assume that $\omega$ is not Martin-L\"of random.
Then a Martin-L\"of test $U_n$, $n=1,2,\dots$, exists such that
$\omega\in\cap U_n$.
Define a Solovay test of randomness as follows. Since $U_n$ is effectively open
and $B_{1/2}(U_n)\le 2^{-n}$ for all $n$, we can effectively compute
a prefix-free sequence of strings $x_n$, $n=1,2,\dots$, such that
$\cup_n\Gamma_{x_n}=\cup_n U_n$ and the series
$\sum\limits_{n=1}^\infty 2^{-l(x_n)}$ converges.
Evidently $x_n\subset\omega$ for infinitely many $n$.

On the other side, assume that for some Solovay test $x_n$, $n=1,2,\dots$,
$x_n\subset\omega$ for infinitely many $n$. Let
$\sum\limits_{n=1}^\infty 2^{-l(x_n)}<2^K$, where $m$ is a positive
integer number. Let $U_n$ be a set of all infinite $\omega$ such that
$|\{m:x_m\subset\omega\}|\ge 2^{n+K}$. It is easy to verify that $U_n$
is a Martin-L\"of test of randomness and that $\omega\in\cap U_n$.
$\triangle$

We also consider total Solovay tests of randomness
which leads to the same definition of randomness as Schnorr tests of
randomness (see~Downey and Griffiths~\cite{DoG2002}).
A series $\sum\limits_{i=1}^\infty r_i$ converges with a computable rate
of convergence if a computable function $m(\delta)$ exists such that
$|\sum\limits_{i=m(\delta)}^\infty r_i|\le\delta$
for each positive rational number $\delta$.
A Solovay test of randomness ${\cal T}=\{x_n:n=1,2,\dots\}$ is called total
if the series $\sum\limits_{n=1}^\infty 2^{-l(x_n)}$ converges with
a computable rate of convergence.
\begin{proposition}\label{Sol-Mart-11}
An infinite sequence $\omega=\omega_1\omega_2\dots$ is Schnorr random
if and only if it passes each total Solovay test of randomness.
\end{proposition}
The proof is similar to the proof of Proposition~\ref{Sol-Mart-1}.

The equivalent definitions of Martin-L\"of random sequence is obtained
in terms of algorithmic complexity (see Li and Vitanyi~\cite{LiV97}).

In terms of prefix complexity the following definition is known.
An infinite sequence $\omega$ is Martin-L\"of random with
respect to a computable measure $P$ if and only if
$
KP(\omega^n)\ge -\log P(\omega^n)+O(1)
$

Analogous definition can be obtained in terms of monotonic complexity.
Let us define a notion of a monotonic computable transformation of
binary sequences. A computable representation of such operation is
a set $\hat\psi\subseteq \{0,1\}^*\times \{0,1\}^*$ such that
(i) the set $\hat\psi$ is recursively enumerable;
(ii) for any $(x,y), (x',y')\in\hat\psi$ if $x\subseteq x'$ then
$y\subseteq y'$ or $y'\subseteq y$;
(iii) if $(x,y)\in\hat\psi$ then $(x,y')\in\hat\psi$ for all $y'\subseteq y$.

The set $\hat\psi$ defines a monotonic with respect to $\subseteq$
decoding function
\footnote
{
Here the by supremum we mean an union of all comparable $x$ in one sequence.
}
$
\psi(p)=\sup\{x:\exists p'(p'\subseteq p\&(p',x)\in\hat{\psi})\}.
$

Any computable monotonic function $\psi$ determines the
corresponding measure of complexity
$
Km_\psi(x)=\min\{l(p):x\subseteq\psi(p)\}=\min\{l(p):(x,p)\in\hat\psi\}.
$
The invariance property also holds for monotonic measures of
complexity: an optimal computable operation $\psi$ exists such
that $Km_\psi(x)\le Km_{\psi'}(x)+O(1)$ for all computable
operations $\psi'$ and for all finite binary sequences $x$.

An infinite sequence $\omega$ is Martin-L\"of random with
respect to a computable measure $P$ if and only if
$
Km(\omega^n)=-\log P(\omega^n)+O(1).
$
In particular, an infinite binary sequence
$\omega$ is Martin-L\"of random
(with respect to uniform measure) if and only if
$Km(\omega^n)=n+O(1)$ (see for details Li and Vitanyi~\cite{LiV97}).

The function
$
dm_P(\omega^n)=-\log P(\omega^n)-Km(\omega^n)
$
is called universal
deficiency of randomness (with respect to a measure $P$). For the
uniform measure, $dm(\omega^n)=n-Km(\omega^n)$.

\section{Algorithmically stable laws}\label{stab-1}

Let $\Phi(\omega)$ be an asymptotic probability law, i.e., a property
of infinite binary sequences which holds almost surely.

Kolmogorov's algorithmic approach to probability
theory offers a new paradigm for logic of probability.
We can formulate an equivalent form of a probabilistic law:
$Km(\omega^n)\ge n-O(1)$ $\Longrightarrow$ $\Phi(\omega)$.

In this paper we present a more deep analysis. We call a law
$\Phi(\omega)$ stable if there exists a unbounded nondecreasing
computable function $\alpha(n)$ such that
$Km(\omega^n)\ge n-\alpha(n)-O(1)$ $\Longrightarrow$ $\Phi(\omega)$.
The function $\alpha(n)$ is called degree of stability of the law
$\Phi(\omega)$.

\subsection{A sufficient condition of stability}

We present in this section a sufficient condition of stability
of a probability law and present examples of probability laws which are
stable with different degree of stability. We formulate this sufficient condition
in terms of Schnorr~\cite{Sch71} definition of algorithmic random sequence.
The choice of Schnorr's definition is justified by an observation that
a vast majority of such laws holds for Schnorr random sequences.

An algorithmic effective version of convergence almost surely
of functions $f_n$ of type $\Omega\to{\cal R}^+$ was introduced
in V'yugin~\cite{Vyu97}. A sequence of functions $f_n$
effectively converges to a function $f$ almost surely if a
computable function $m(\delta,\epsilon)$ exists such that
\begin{eqnarray}
B_{1/2}\{\omega:\sup\limits_{n\ge m(\delta,\epsilon)}
|f_n(\omega)-f(\omega)|>\delta\}<\epsilon
\label{eff-conv-1}
\end{eqnarray}
for all positive rational numbers $\delta$ and $\epsilon$.

The following simple proposition was formulated in~\cite{Vyu97} for
Martin-L\"of notion of randomness. It holds also for Schnorr random sequences.
\begin{proposition}\label{conf-1}
Let a computable sequence of functions $f_n$ effectively converges
almost surely to some function $f$. Then a Schnorr test of randomness
$\cal T$ can be constructed such that
$\lim\limits_{n\to\infty}f_n(\omega)=f(\omega)$ for each infinite sequence
$\omega$ passing the test $\cal T$.
\end{proposition}
{\it Proof}. By (\ref{eff-conv-1}) we have
$
B_{1/2}\{\omega:\sup\limits_{n,n'\ge m(\delta/2,\epsilon)}
|f_n(\omega)-f_{n'}(\omega)|>\delta\}<\epsilon
$
for all positive rational numbers $\delta$ and $\epsilon$.
Denote $W_{n,n',\delta}=\{\omega:|f_n(\omega)-f_{n'}(\omega)|>\delta\}$.
This set can be represented as the union $\cup_i\Gamma_{x_i}$,
where $x_i$, $i=1,2,\dots$, is a computable sequence of finite sequences.
Define $V_i=\cup_{n,n'\ge m(1/i,2^{-i})}W_{n,n',1/i}$ for all $i$
and $U_i=\cup_{j>i} V_j$. Then $B_{1/2}(U_i)\le 2^{-i}$ for all $i$.

Note that the measure $B_{1/2}(U_i)$ can be computed with an arbitrary degree
of precision. Indeed by (\ref{eff-conv-1}), to calculate $P(U_i)$
with a given degree of precision $\epsilon>0$ it is sufficient to calculate
$B_{1/2}(\cup_{i'\ge j\ge i}\cup_{m'\ge n,n'\ge m(1/i,2^{-j})}W_{n,n',1/j})$ for some sufficiently
large $i'$ and $m'$. Therefore, ${\cal T}=\{U_i\}$ is the Schnorr
test of randomness.

Assume that $\lim\limits_{n\to\infty}f_n(\omega)$ does not exist for some
$\omega$. Then an $i$ exists such that $|f_n(\omega)-f_{n'}(\omega)|>1/i$
for infinitely many $n$ and $n'$. For any $j>i$ the numbers
$n,n'\ge m(1/j,2^{-j})$ exist such that $\omega\in W_{n,n',1/j}\subseteq V_j$.
Hence, the sequence $\omega$ if rejected by the Schnorr test
${\cal T}$. $\triangle$

In the following theorem some sufficient condition of stability of
a probability law is given in terms of total Solovay tests randomness.

\begin{proposition}\label{suff-1}
For any total Solovay test of randomness
${\cal T}$, a computable unbounded function $\sigma(n)$ exists
such that for any infinite sequence $\omega$ if
$Km(\omega^n)\ge n-\sigma(n)-O(1)$ then the sequence $\omega$
passes the test ${\cal T}$.
\end{proposition}
{\it Proof}. Let ${\cal T}=\{x_n:n=1,2,\dots\}$. Denote $l_s=l(x_s)$.
Since
$
\sum\limits_{s=1}^\infty 2^{-l_s}<\infty
$
with a uniform computable rate of convergence $m(\epsilon)$, an unbounded nondecreasing
computable function $\nu(n)$ exists such that
$
\sum\limits_{s=1}^\infty 2^{-l_s+\nu(l_s)}<\infty.
$
We can define $\nu(n)=i$, where $i$ is such that
$m(2^{-2i})\le n<m(2^{-2(i+1)})$.
Then
$$
\sum\limits_{s=1}^\infty 2^{-l_s+\nu(l_s)}=
\sum\limits_{i=1}^\infty 2^i
\sum\limits_{m(2^{-2i})\le l_s<m(2^{-2(i+1)})} 2^{-l_s}\le
\sum\limits_{i=1}^\infty 2^{-i}\le 1.
$$
By the generalized Kraft inequality (see Li and Vitanyi~\cite{LiV97}),
we can define the corresponding prefix-free code such that
$
Km(x_m)\le l(x_m)-\nu(l(x_m))+O(1)
$
Assume $x_m\subseteq\omega$ for
infinitely many $m$. For any such $m$, $\omega^n=x_m$, where $n=l(x_m)$.

Let $\sigma(n)$ be a unbounded nondecreasing computable function such that
$\sigma(n)=o(\nu(n))$ as $n\to\infty$. Let also, $\omega$ be an infinite
binary sequence such that $Km(\omega^n)\ge n-\sigma(n)-O(1)$ for all $n$.
For $n=l(x_m)$,
\begin{eqnarray*}
\sigma(n)\ge n-Km(\omega^n)\ge
n-l(x_m)+\nu(l(x_m))-O(1)
\ge
\nu(n)-O(1)
\end{eqnarray*}
for infinitely many $n$. On the other hand,
$\sigma(n)=o(\nu(n))$ as $n\to\infty$.
This contradiction proves the theorem.
$\triangle$

By Proposition~\ref{suff-1} stability property holds for main probability laws
like the strong law of large numbers and the law of iterated logarithm.

By computable sequence of total Solovay tests of randomness we mean
a computable double indexed sequence of finite binary
strings ${\cal T}_k=\{x_{k,n}:n=1,2,\dots\}$, $k=1,2,\dots$, such that the series
$\sum\limits_{n=1}^\infty 2^{-l(x_{k,n})}$
converges with a uniformly by $k$ computable rate of convergence.
This means that
there exists a computable function
$m(\delta,k)$ such that
$\sum\limits_{i=m(\delta,k)}^\infty 2^{-l(x_{k,i})}\le\delta$ for each $k$
and rational $\delta$.
\footnote{
We can combine all tests of computable sequence ${\cal T}_k$,
$k=1,2,\dots$, in a single total test
${\cal T}=\{x_{k,n}:k=1,2,\dots, n=m(2^{-k},k),m(2^{-k},k)+1,\dots\}$
such that if any $\omega$ passes the test ${\cal T}$ then it passes the test
${\cal T}_k$ for each $k$. ${\cal T}$ is the test, since
$\sum\limits_{k=1}^\infty\sum\limits_{n=m(2^{-k},k)}^\infty 2^{-l(x_{k,n})}\le
\sum\limits_{k=1}^\infty 2^{-k}\le 1$.
}

In applications, often convenient to use computable sequences of tests.
Easy to modify Proposition~\ref{suff-1} for computable sequences of tests.
\begin{proposition}\label{suff-2}
For any computable sequence of Solovay total tests of randomness
${\cal T}_k$, $k=1,2,\dots$, a computable unbounded function $\sigma(n)$ exists
such that for any infinite sequence $\omega$ if
$Km(\omega^n)\ge n-\sigma(n)-O(1)$ then the sequence $\omega$
passes all tests ${\cal T}_k$.
\end{proposition}
The proof is analogous to the proof of Proposition~\ref{suff-1}.

Let us show that the strong law of large numbers
corresponds to the computable sequence of total Solovay randomness tests.

Hoeffding~\cite{Hoe63} inequality for uniform probability distribution
\begin{eqnarray}
B_{1/2}\left\{\omega\in\Omega: \left|\frac{1}{n}\sum\limits_{i=1}^n
\omega_i-\frac{1}{2}\right|\ge\epsilon\right\}\le 2e^{-2n\epsilon^2}
\label{strong-1g}
\end{eqnarray}
serves as a tool for constructing total Solovay tests of randomness.

Let $\epsilon_k$ be a computable sequence of positive rational numbers
such that $\epsilon_k\to 0$ as $k\to\infty$. For any $k$, let
$\cup_n\{x:l(x)=n\&|\frac{1}{n}\sum\limits_{i=1}^n
x_i-\frac{1}{2}|\ge\epsilon_k\}=\{x_{k,m}:m=1,2,\dots\}\}$.

This is the total Solovay tests of randomness, since by (\ref{strong-1g})
it holds
$
\sum\limits_{m=1}^\infty 2^{-l(x_{k,m})}\le
\sum\limits_{n=1}^\infty 2e^{-2n\epsilon^2_k}<\infty
$
with a computable rate of convergence.

The strong law of large numbers
$
\lim\limits_{n\to\infty}\frac{1}{n}\sum\limits_{i=1}^n\omega_i=\frac{1}{2}
$
holds for an infinite sequence $\omega=\omega_1\omega_2\dots$
if and only if it passes the test $\{x_{k,m}:m=1,2,\dots\}$ for each $k$.
By Proposition~\ref{suff-1} an unbounded nondecreasing
computable function $\sigma(n)$ exists
such that if $Km(\omega^n)\ge n-\sigma(n)-O(1)$ as $n\to\infty$ then
the strong law of large numbers holds for this $\omega$.

We can find the specific form of this function $\sigma(n)$
using the proof of Proposition~\ref{suff-1}. By the inequality
(\ref{strong-1g}) we have the bound $\sum\limits_{n=1}^\infty
2^{-l(x_{k,n})}< \sum\limits_{n=1}^\infty
2e^{-2n\epsilon^2_k}<\infty$ for the corresponding total
Solovay test of randomness ${\cal T}_k=\{x_{k,n}\}$. Also,
$\sum\limits_{n=1}^\infty 2e^{-2n\epsilon^2_k+\nu(n)}<\infty$
for any function $\nu(n)$ such that $\nu(n)=o(n)$ as
$n\to\infty$. The rest part of the proof coincides with the
proof of Proposition~\ref{suff-1}. Hence, any function
$\sigma(n)=o(n)$ can serve as a degree of stability for the
strong law of large numbers.

An analogous construction can be developed for the law of iterated logarithm:
\begin{eqnarray}
\limsup\limits_{n\to\infty}\frac{\left|\sum\limits_{i=1}^n\omega_i-\frac{n}{2}\right|}
{\sqrt{\frac{1}{2}n\ln\ln n}}=1.
\label{iterr-1}
\end{eqnarray}
We consider here only the inequality $\le$ in (\ref{iterr-1}).
\footnote{
The converse inequality is studied in Vovk~\cite{Vov87}.
}
This inequality violates if and only if a rational number
$\delta>1$ exists such that
$
S_n-\frac{n}{2}>\delta\sqrt{\frac{1}{2}n\ln\ln n}
$
for infinitely many $n$, where $S_n=\sum\limits_{i=1}^n\omega_i$.

For any rational number $\delta$ such that $\delta>1$ and for
$m_{n}=\lceil\delta^{n}\rceil$, let
\footnote{For any real number $r$, $\lceil r\rceil$ denotes the least
positive integer number $m$ such that $m\ge r$.}

\begin{eqnarray}
U_{\delta,n}=
\{\omega\in\Omega: \exists k(m_{n}\le k\le m_{n+1}\&
S_k-k/2>\delta\sqrt{(1/2)m_n\ln\ln m_n}\}.
\nonumber
\end{eqnarray}
Using the inequality
$
B_{1/2}\{\max\limits_{1\le k\le m} S_k>a\}\le 2 B_{1/2}\{S_m>a\},
$
we obtain
\begin{eqnarray}
B_{1/2}(U_{\delta,n})\le
2 B_{1/2}(\{\omega\in\Omega: S_{m_{n+1}}-m_{n+1}/2>
\delta\sqrt{(1/2)m_{n}\ln\ln m_{n}}\})\le
\nonumber
\\
\le ce^{-\delta\ln\ln m_{n}}\approx\frac{1}{n^\delta},~~~~~
\label{iterr-3}
\end{eqnarray}
where $c>0$. We have used in (\ref{iterr-3}) the
Hoeffding inequality.

We can effectively construct a prefix-free set $\tilde U_{\delta,n}$
of finite sequences such that for each $\omega\in U_{\delta,n}$
an $m$ exists such that $\omega^m\in\tilde U_{\delta,n}$.

A sequence
$\cup_n\tilde U_{\delta,n}=\{x_{\delta,k}:k=1,2,\dots\}$
is a total Solovay test of randomness, since the series
$
\sum\limits_n 2^{-l(x_{\delta,n})}=\sum\limits_n B_{1/2}(U_{\delta,n})\le
\sum\limits_n \frac{1}{n^\delta}
$
converges (with a computable rate of convergence) for any $\delta>1$.

By definition the law of iterated logarithm (\ref{iterr-1}) holds
for $\omega=\omega_1\omega_2\dots$
if and only if it passes the test $\{x_{\delta,k}:k=1,2,\dots\}$
for each $\delta>1$.

By Proposition~\ref{suff-1} an unbounded nondecreasing
computable function $\sigma(m)$ exists such that the inequality $\le$ in
(\ref{iterr-1}) holds
for any $\omega$ satisfying $Km(\omega^m)\ge m-\sigma(m)-O(1)$ as $m\to\infty$.

We can also find a specific form of the degree of stability for the law of iterated logarithm.
Let $\alpha(m)$ be a unbounded nondecreasing computable
function such that $\alpha(m)=o(\ln\ln m)$ as $m\to\infty$. Then the series
$
\sum\limits_n e^{-\delta\ln\ln m_{n}+\alpha(m_n)}\approx
\sum\limits_n\frac{o(\ln n)}{n^\delta}
$
converges for any $\delta>1$.
The proof of Proposition~\ref{suff-1} shows that any computable unbounded
function $\sigma(n)=o(\log\log n)$ can serve as a measure of stability
of the law of iterated logarithm.

\subsection{Stability of the Birkhoff's theorem in ergodic case}

Recall some basic notions of ergodic theory.
An arbitrary measurable mapping of the a probability space
into itself is called a transformation.
A transformation $T:\Omega\to\Omega$ preserves a measure $P$ on $\Omega$ if
$P(T^{-1}(A))=T(A)$ for all measurable subsets $A$ of the space.
A subset $A$ is called invariant with respect to $T$ if $T^{-1}A=A$ up to
a set of measure 0.
A transformation $T$ is called ergodic if each invariant with respect
to $T$ subset $A$ has measure~0~or~1.

A transformation $T$ of the set $\Omega$ is computable if a computable
representation $\hat\psi$ exists such that (i)-(iii) hold and
$
T(\omega)=\sup\{y:x\subseteq\omega\&(x,y)\in\hat{\psi})\}
$
for all infinite $\omega\in\Omega$.


Denote $T^0\omega=\omega$, $T^{i+1}\omega=T(T^i\omega)$.
Any point $\omega\in\Omega$ generates an infinite
trajectory $\omega, T\omega,T^2\omega,\dots$.

Using Bishop's~\cite{Bis67} analysis, V'yugin~\cite{Vyu97},~\cite{Vyu98}
presented an algorithmic version of the Birkhoff's pointwise
ergodic theorem:

Let $T$ be a computable measure preserving transformation and
$f$ be a computable real-valued bounded
function defined on the set of binary sequences. Then
for any infinite binary sequence $\omega$ the following implication is valid:
\begin{eqnarray}
Km(\omega^n)\ge n-O(1)\Longrightarrow\lim\limits_{n\to\infty}\frac{1}{n}
\sum\limits_{i=0}^{n-1}f(T^i\omega)=\hat f(\omega)
\label{e-1}
\end{eqnarray}
for some $\hat f(\omega)$ ($=E(f)$ for ergodic $T$).

Later this result was extended for non-computable $f$ and generalized
for more general metric spaces.
For further development see Nandakumar~\cite{Nan2008},
Galatolo et al.~\cite{GHR2010}, and Gacs et al.~\cite{GHR2011}.

Let $f\in L^1$ be computable and 
$\sup_\omega |f(\omega)|<\infty$, $P$ be a computable measure and $T$ be
a computable ergodic transformation preserving the measure $P$.
By $\|f\|$ denote the norm in $L^1$ (or in $L^2$).

Define the sequence of ergodic averages $A^f_n$, $n=1,2,\dots$, where
$A^f_n(\omega)=\frac{1}{n}\sum\limits_{k=0}^{n-1}f(T^k\omega)$.

Galatolo at al.~\cite{GHR2010a} show that
ergodic averages $\{A^f_n\}$ effectively converges to some
computable real number $c=\int f(\omega)dP$ almost surely as $n\to\infty$.
Then the stability property of the ergodic theorem in case where the
transformation $T$ is ergodic is the corollary of this result and
Propositions~\ref{conf-1}~and~\ref{suff-1}.
We present this result for completeness of exposition.
\begin{proposition}\label{erg-aver-1a}
The sequence of ergodic averages $\{A^f_n\}$ effectively converges 
almost surely as $n\to\infty$.
\end{proposition}
{\it Proof}. At first we prove effective convergence in norm $L^1$
and thereafter we will use the maximal ergodic theorem.

We suppose without loss of generality that $\int fdP=0$.
\footnote{
Replace $f$ on $f-\int f(\omega)dP$.
}
Then $\|A^f_n\|\to 0$ as $n\to\infty$.

Given $\epsilon>0$ compute an $p(\epsilon)$ such that $\|A^f_p\|<\epsilon/2$
for $p=p(\epsilon)$. Let $m=np+k$, where $0\le k<p$. Then
$A^f_m(\omega)=\frac{1}{m}\left(\sum\limits_{i=0}^{n-1}pA^f_p(T^{pi}\omega)
+kA^f_k(T^{pn}\omega)\right)$. Then $\|A^f_m\|\le\frac{1}{m}(np\|A^f_p\|+
k\|A^f_k\|)\le
\|A^f_p\|+\frac{1}{n}\|A^f_k\|\le\|A_p\|+\frac{1}{n}\|f\|<\epsilon/2+
\frac{1}{n}\|f\|\le\epsilon$ for all
$n\ge n(\epsilon)=\max\{p(\epsilon),(2/\epsilon)\|f\|\}$.

The maximal ergodic theorem says that
$B_{1/2}\{\omega:\sup\limits_n|A^f_n(\omega)|>\delta\}\le
\frac{1}{\delta}\|f\|$ for any ergodic transformation $T$ preserving measure
$P$. Given $\epsilon,\delta>0$, compute an
$p=p(\delta,\epsilon)$ such that
$\|A^f_p\|\le\delta\epsilon/2$. By the maximal ergodic theorem for $g=A^f_p$
we have $B_{1/2}\{\omega:\sup\limits_n |A^g_n(\omega)|>\delta/2\}\le\epsilon$.

Now, we check that $A^g_n$ is not too far from $A_n^f$. Indeed,
\begin{eqnarray*}
A_n^g(\omega)=\frac{1}{n}\sum\limits_{k=0}^{n-1}g(T^k\omega)=
\frac{1}{np}\sum\limits_{k=0}^{p-1}\sum\limits_{s=0}^{n-1}f(T^{k+s}\omega)=
\frac{1}{np}\left(p\sum\limits_{k=0}^{n-1}f(T^k\omega)\right)+
\\
+\frac{1}{np}\left(\sum\limits_{k=0}^{p-1}(p-k)f(T^k(T^n\omega))\omega)-
\sum\limits_{k=0}^{p-1}(p-k)f(T^k\omega)\right).
\end{eqnarray*}
This implies that $\sup\limits_\omega |A_n^g(\omega)-A_n^f(\omega)|\le
\frac{2}{np}\sum\limits_{k=0}^{p-1}(p-k)\sup\limits_\omega|f(\omega)|=
\frac{p-1}{n}\sup\limits_\omega|f(\omega)|\le\delta/2$ for all
$n\ge m(\delta,\epsilon)=2(p(\delta,\epsilon)-1)\sup\limits_\omega |f(\omega)|/\delta$.
Hence,
$B_{1/2}\{\omega:\sum\limits_{n\ge m(\delta,\epsilon)}|A_n^f(\omega)|>\delta\}\le\epsilon$.
Proposition is proved. $\triangle$

Propositions~\ref{conf-1},~\ref{suff-1}, and~\ref{erg-aver-1a} imply
a stable version of the ergodic theorem in case where the transformation $T$
is ergodic.
\begin{theorem}\label{stability-1d}
Let $f$ be a computable observable, $T$ be a computable ergodic transformation
preserving the uniform measure $B_{1/2}$. Then a computable unbounded nondecreasing
function $\sigma(n)$ exists such that for any infinite sequence $\omega$
the condition
$Km(\omega^n)\ge n-\sigma(n)-O(1)$ implies that the limit
$\lim\limits_{n\to\infty}\frac{1}{n}\sum\limits_{k=0}^{n-1}f(T^k\omega)$
exists.
\end{theorem}
In particular, in case where transformation $T$ is ergodic,
the ergodic theorem holds for any Schnorr random sequence.
\footnote{This folkloric result was first published by
Franklin and Towsner~\cite{FrT14}.}

\section{Instability in ergodic theory}\label{instab-1}

The phenomenon of instability occurs in ergodic theory.
In this section we present property of uniform instability
of the ergodic theorem and absolute instability for non-ergodic transformation.

\subsection{Instability results}

The degree of stability $\sigma(n)$ from Theorem~\ref{stability-1d} may
depend on observable $f$ and transformation $T$. The following
Theorem~\ref{th-1} shows that there is no uniform degree on stability
$\sigma(n)$ for the ergodic theorem.

Phenomenon of instability of the ergodic theorem was first discovered in
V'yugin~\cite{Vyu2001}.
Compared with ``symbolic dynamics type'' result from~\cite{Vyu2001},
this result is ``measure free'' -- it
is formulated in terms of transformations and Kolmogorov complexity. 
\begin{theorem}\label{th-1}
Let $\sigma(n)$ be a nondecreasing unbounded computable function. Then there
exist a computable ergodic measure preserving transformation $T$
and an infinite sequence $\omega\in\Omega$ such that the inequality
$Km(\omega^n)\ge n-\sigma(n)$ holds for all $n$ and the limit
\begin{eqnarray}
\lim\limits_{n\to\infty}\frac{1}{n}
\sum\limits_{i=0}^{n-1}f(T^i\omega)
\label{e-1a}
\end{eqnarray}
does not exist for some computable indicator function $f$.
\end{theorem}
In the next theorem an uniform with respect to $\sigma(n)$
result is presented. In this case, we will lose the ergodic
property of transformation $T$.
\begin{theorem}\label{th-1b}
A computable measure preserving transformation $T$ can be constructed such that
for any nondecreasing unbounded computable function $\sigma(n)$
an infinte sequence $\omega$ exists such that $Km(\omega^n)\ge n-\sigma(n)$
holds for all $n$ and the limit (\ref{e-1a}) does not exist for some
computable indicator function $f$.
\end{theorem}
The construction of the transformation $T$ is given in Section~\ref{sec-1};
the proof of Theorem~\ref{th-1} is given in Section~\ref{sec-2}.
In Section~\ref{cat-sta} we consider the main technical concept -- the method
cutting and stacking.

\subsection{Method of cutting and stacking}\label{app}
\label{cat-sta}

In this section we consider the main notions and properties of cutting
and stacking method (see Shields~\cite{Shi91, Shi93}).

A column is a sequence $E=(L_1,\dots,L_h)$ of pairwise disjoint
intervals of the unit interval $[0,1]$ of equal width:
$L_1,\dots, L_h$. We refer to $L_1$ as to the base and to $L_h$ as to
the top of the column; ${\hat E}=\cup_{i=1}^{h}L_i$ is the support of the
column, $w(E)=\lambda(L_1)$ is the width of the column,
$h$ is the height of the column,
$\lambda({\hat E})=\lambda(\cup_{i=1}^{h}L_i)$ is the measure of the column,
where $\lambda$ is uniform measure in $[0,1]$.

Any column defines a transformation $T$
which linearly transforms $L_j$ to $L_{j+1}$,
namely, $T(x)=x+c$ for all $x\in L_j$, where $c$ is the corresponding constant.
This transformation $T$ is not defined outside all intervals of the column
and at all points of the top $L_h$ interval of this column.

Denote $T^0\omega=\omega$, $T^{i+1}\omega=T(T^i\omega)$.
For any $1\le j<h$, an arbitrary point $\omega\in L_j$ generates a finite
trajectory $\omega, T\omega,\dots, T^{h-j}\omega$.

A partition $\pi=(\pi_0,\pi_1)$ is compatible with a column $E$
if for each $j$ there exists an $i$ such that $L_j\subseteq\pi_i$.
This number $i$ is called the name of the interval $L_j$, and the
corresponding sequence of names of all intervals of the column is
called the name of the column $E$.

For any point $x\in L_j$, where $1\le j<h$, by $E$--name of
the trajectory $x, Tx,\dots, T^{h-j}x$ we mean
a sequence of names of intervals $L_j,\dots, L_h$ from the column $E$.
The length of this sequence is $h-j+1$.

A gadget is a finite collection of disjoint columns.
The width of the gadget $w(\Upsilon)$ is the sum of the widths of its
columns. A union of gadgets $\Upsilon_i$ with disjoint supports
is the gadget $\Upsilon=\cup\Upsilon_i$ whose columns are the columns
of all the $\Upsilon_i$. The support of the gadget $\Upsilon$ is
the union $\hat\Upsilon$ of the supports of all its columns.
A transformation $T=T(\Upsilon)$ is associated with a gadget $\Upsilon$ if
it is the union of transformations defined on all columns of $\Upsilon$.
With any gadget $\Upsilon$ the corresponding set of finite trajectories
generated by points of its columns is associated. By $\Upsilon$-name
of a trajectory we mean its $E$-name, where $E$ is that column
of $\Upsilon$ to which this trajectory corresponds.
A gadget $\Upsilon$ extends a column $\Lambda$ if the support of
$\Upsilon$ extends the support of $\Lambda$, the transformation
$T(\Upsilon)$ extends the transformation $T(\Lambda)$ and the partition
corresponding to $\Upsilon$ extends the partition corresponding to $\Lambda$.

Since all points of the interval $L_j$ of the column generate
identical trajectories,
we refer to this trajectory as to the trajectory generated by the
interval $L_j$.

The cutting and stacking operations that are common used will now be
defined. The distribution of a gadget $\Upsilon$ with columns
$E_1,\dots,E_n$ is a vector of probabilities
\begin{eqnarray}
\left(\frac{w(E_1)}{w(\Upsilon)},\dots,\frac{w(E_n)}{w(\Upsilon)}\right).
\label{gad-division}
\end{eqnarray}
A gadget $\Upsilon$ is a copy of a gadget $\Lambda$ if they have the same
distributions and the corresponding columns have the same
partition names.

A gadget $\Upsilon$ can be cut into $M$ copies of itself
$\Upsilon_m, m=1,\dots, M$, according to a given probability vector
$(\gamma_1,\dots,\gamma_M)$ of type (\ref{gad-division}) by cutting each column
$E_i=(L_{i,j}: 1\le j\le h(E_i))$ (and its intervals) into disjoint
subcolumns $E_{i,m}=(L_{i,j,m}: 1\le j\le h(E_i))$ such that
$w(E_{i,m})=w(L_{i,j,m})=\gamma_m w(L_{i,j})$.

The gadget $\Upsilon_m=\{E_{i,m}:1\le i\le L\}$ is called the copy of
the gadget $\Upsilon$ of width $\gamma_m$. The action of the gadget
transformation $T$ is not affected by the copying operation.

Another operation is the stacking gadgets onto gadgets.
At first we consider the stacking of columns onto columns and
the stacking of gadgets onto columns.

Let $E_1=(L_{1,j}:1\le j\le h(E_1))$ and $E_2=(L_{2,j}:1\le j\le h(E_2))$
be two columns of equal width whose supports are disjoint.
The new column $E_1*E_2=(L_j:1\le j\le h(E_1)+h(E_2))$ is defined as
$L_j=L_{1,j}$ for all $1\le j\le h(E_1)$ and $L_j=L_{2,j-h(E_1)+1}$ for all
$h(E_1)\le j\le h(E_1)+h(E_2)$.

Let a gadget $\Upsilon$ and a column $E$ have the same width, and their
supports are disjoint. A new gadget $E*\Upsilon$ is defined as follows.
Cut $E$ into subcolumns $E_i$ according to the distribution of the gadget
$\Upsilon$ such that $w(E_i)=w(U_i)$, where $U_i$ is the $i$th column
of the gadget $\Upsilon$. Stack $U_i$ on the top of $E_i$ to get the new
column $E_i*U_i$. A new gadget consists of the columns $(E_i*U_i)$.

Let $\Upsilon$ and $\Lambda$ be two gadgets of the same width
and with disjoint supports. A gadget $\Upsilon*\Lambda$ is defined as
follows. Let the columns of $\Upsilon$ are $\{E_i\}$. Cut $\Lambda$ into
copies $\Lambda_i$ such that $w(\Lambda_i)=w(E_i)$ for all $i$.
After that, for each $i$ stack the gadget $\Lambda_i$ onto
column $E_i$, ie, we consider a gadget $E_i*\Lambda_i$.
The new gadget is the union of gadgets $E_i*\Lambda_i$ for all $i$.
The number of columns of the gadget $\Upsilon*\Lambda$ is the product
of the number of columns of $\Upsilon$ on the number of columns of
$\Lambda$.

The $M$-fold independent cutting and stacking of a single gadget
$\Upsilon$ is defined by cutting $\Upsilon$ into $M$ copies $\Upsilon_i$,
$i=1,\dots,M$, of equal width and successively independently cutting
and stacking them to obtain $\Upsilon^{*(M)}=\Upsilon_1*\dots*\Upsilon_M$.
A sequence of gadgets $\{\Upsilon_m\}$ is complete if
\begin{itemize}
\item{}
$\lim\limits_{m\to\infty} w(\Upsilon_m)=0$;
\item{}
$\lim\limits_{m\to\infty} \lambda({\hat\Upsilon}_m)=1$;
\item{}
$\Upsilon_{m+1}$ extends $\Upsilon_m$ for all $m$.
\end{itemize}
Any complete sequence of gadgets $\{\Upsilon_s\}$ determines a transformation
$T=T\{\Upsilon_s\}$ which is defined almost surely.

By definition $T$ preserves the measure $\lambda$. In Shields~\cite{Shi91}
the conditions sufficient a process $T$ to be ergodic
were suggested. Let a gadget $\Upsilon$ is constructed by cutting and
stacking from a gadget $\Lambda$.
Let $E$ be a column from $\Upsilon$ and $D$ be a column from $\Lambda$.
Then ${\hat E}\cap{\hat D}$ is defined as the union of subcolumns from
$D$ of width $w(E)$ which were used for construction of $E$.

Several examples of stationary measures constructed using cutting and stacking
method are given in Shields~\cite{Shi91, Shi93}. We use in
Section~\ref{sec-2}
a construction of a sequence of gadgets defining the uniform Bernoulli
distribution on trajectories generated by them. This sequence
is constructed using the following scheme. Let a partition
$\pi=(\pi_0,\pi_1)$ be given. Let also $\Delta$ be a gadget such that
its columns have the same width and are compatible with the partition
$\pi$. Let $\lambda(\hat\Delta\cap\pi_0)=\lambda(\hat\Delta\cap\pi_1)$.
Suppose that for some $M$ a gadget $\Delta'$ is constructed
from the gadget $\Delta$ by means of $M$-fold independent cutting
and stacking.
Then $B_{1/2}(x_1\dots x_n)=2^{-n}\lambda(\hat\Delta)$ for the trajectory
$x_1\dots x_n$ of any point of the support of $\hat\Delta'$.

Let $0<\epsilon<1$.
A gadget $\Lambda$ is $(1-\epsilon)$-well-distributed in $\Upsilon$ if
\begin{equation}
\sum_{D\in\Lambda}\sum_{E\in\Upsilon}|\lambda({\hat E}\cap{\hat D})-
\lambda({\hat E})\lambda({\hat D})|<\epsilon.
\end{equation}
We will use the following two lemmas.
\begin{lemma} \label{well-def}
(\cite{Shi91}, Corollary 1), (\cite{Shi93}, Theorem A.1).
Let $\{\Upsilon_n\}$ be a complete sequence of gadgets and for each $n$
the gadget $\{\Upsilon_n\}$ is $(1-\epsilon_n)$-well-distributed in
$\{\Upsilon_{n+1}\}$, where $\epsilon_n\to 0$. Then $\{\Upsilon_n\}$
defines the ergodic process.
\end{lemma}
\begin{lemma} \label{M-fold} (\cite{Shi93}, Lemma 2.2).
For any $\epsilon>0$ and any gadget $\Upsilon$ there is an $M$ such that
for each $m\ge M$ the gadget $\Upsilon$ is $(1-\epsilon)$-well-distributed
in the gadget $\Upsilon^{*(m)}$ constructed from $\Upsilon$ by
$\mbox m$-fold independent cutting and stacking.
\end{lemma}
The proof is given in Shields~\cite{Shi91}.


\subsection{Construction}~\label{sec-1}

Let $r>0$ be a sufficiently small rational number. Define a partition
$\pi=(\pi_0,\pi_1)$ of the unit interval $[0,1]$, where
$\pi_0=[0,0.5)\cup (0.5+r,1)$ and $\pi_1=[0.5,0.5+r]$.

Let $\sigma(n)$ be a computable unbounded nondecreasing function.
A computable sequence of positive integer numbers
exists such that $0<h_{-2}<h_{-1}<h_0<h_1<\dots$ and
$
\sigma(h_{i-1})-\sigma(h_{i-2})>i-\log r+11
$
for all $i=0,1,\dots$.

The gadgets $\Delta_{s}$, $\Pi_{s}$, where $s=0,1,\dots$,
will be defined by mathematical induction on steps.
The gadget $\Delta_{0}$ is defined by cutting of the interval
$[0.5-r,0.5+r]$ on equal parts and by stacking them.
Let $\Pi_0$ be a gadget defined by cutting of the intervals
$[0,0.5-r)$ and $(0.5+r,1]$ in equal subintervals
and stacking them. The purpose of this definition is to construct initial
gadgets of width $\le 2^{-h_0}$ with supports satisfying
$\lambda(\hat\Delta_0)=2r$ and $\lambda(\hat\Pi_0)=1-2r$.

The sequence of gadgets $\{\Delta_{s}\}$, $s=0,1,\dots$, will
define an approximation of the uniform Bernoulli measure
concentrated on the names ot their trajectories (see Section~\ref{cat-sta}).
The sequence of gadgets $\{\Pi_s\}$, $s=0,1,\dots$, will define a measure
with sufficiently small entropy.
The gadget $\Pi_{s-1}$ will be extended at each step of the
construction by the half part of the gadget $\Delta_{s-1}$. After that, the
independent cutting and stacking process will be applied to this extended
gadget. This process eventually defines infinite trajectories starting from
points of $[0,1]$. The sequence of gadgets
$\{\Pi_s\}$, $s=0,1,\dots$, will be complete and will define
a transformation $T$.
Lemmas~\ref{well-def} and \ref{M-fold} from Section~\ref{app} ensure
the transformation $T$ to be ergodic.

{\it Construction}.
Let at step $s-1$ ($s>0$) gadgets $\Delta_{s-1}$ and $\Pi_{s-1}$
were defined. Cut of the gadget $\Delta_{s-1}$ into two copies
$\Delta'$ and $\Delta''$ of equal width (i.e. we cut of each column into two
subcolumns of equal width) and join $\Pi_{s-1}\cup\Delta''$
in one gadget. Find a sufficiently large
number $R_s$ and do $R_s$-fold independent cutting
and stacking of the gadget $\Pi_{s-1}\cup\Delta''$ and also of the
gadget $\Delta'$ to obtain new gadgets $\Pi_s$ of width $\le 2^{-h_s}$
and $\Delta_{s}$ such that the gadget $\Pi_{s-1}\cup\Delta^{''}$ is
$(1-1/s)$--well--distributed in the gadget $\Pi_s$. The needed
number $R_s$ exists by Lemma~\ref{M-fold} (Section~\ref{app}).

By construction, the endpoints of all subintervals of $[0,1]$
used in this construction are rational
numbers, and so, the construction is algorithmically effective.

{\it Properties of the construction}. Define a transformation $T=T\{\Pi_s\}$.
Since the sequence of the gadgets $\{\Pi_s\}$ is complete
(i.e. $\lambda({\hat\Pi}_s)\to 1$ and $w(\Pi_s)\to 0$
as $s\to\infty$), $T$ is defined almost surely.

The transformation $T$ is ergodic by Lemma~\ref{well-def},
since the sequence of gadgets
$\Pi_s$ is complete. Besides, the gadget $\Pi_{s-1}\cup\Delta''$,
and the gadget $\Pi_{s-1}$ are $(1-1/s)$-well distributed
in $\Pi_s$ for any $s$. By construction
$\lambda(\hat\Delta_i)=2^{-i+1}r$ and $\lambda(\hat\Pi_i)=1-2^{-i+1}r$
for all $i=0,1,\dots$.

We need to interpret the transformation $T$ as a transformation of
infinite binary sequences. To do this,
we identify real numbers from $[0,1]$ with their infinite binary
representations. This correspondence in one-to-one besides the countable set
of infinite sequences corresponding to dyadic rational numbers: for example,
$0.0111...=0.10000...$.
\footnote{
Such sequences can be ignored, since they are not Martin-L\"of
random with respect to the uniform Bernoulli measure on $\Omega$.
In particular, the measure of the set of all such sequences is zero.}
From the point of view of this interpretation,
the Bernoulli measure $B_{1/2}$ and the uniform measure $\lambda$
are identical and transformation $T$ constructed above preserves the uniform
Bernoulli measure and is defined almost surely.

\subsection{Proof of Theorem~\ref{th-1}}~\label{sec-2}

For technical convenience, we replace in the proof of Theorem~\ref{th-1}
the deficiency of randomness $dm(x)$ by a notion of supermartingale
(see Schiryaev~\cite{Shi80}).
A function $M:\{0,1\}^*\to{\cal R}$ is called supermartingale if
$M(\Lambda)\leq 1$ and $M(x)\ge\frac{1}{2}(M(x0)+M(x1))$ for all $x$.
Also, we require $M(x)\ge 0$ for all $x$. More general property holds:
$M(x)\ge\sum\limits_{y\in B}M(xy)2^{-l(y)}$ for any prefix-free set $B$.

Let us prove that the deficiency of randomness is bounded by logarithm
of some supermartingale: $dm(x)\le\log M(x)$ for all $x$.

Let the optimal function $\psi$ defines the monotone complexity $Km(x)$. Define
$Q(x)=B_{1/2}(\cup\{\Gamma_p:x\subseteq\psi(p)\})$.
It is easy to verify that
$Q(\Lambda)\le 1$ and $Q(x)\ge Q(x0)+Q(x1)$
for all $x$. Then the function $M(x)=2^{l(x)}Q(x)$
is a supermartingale and $M(x)\ge 2^{-Km(x}$ for all $x$.

Denote $d(x)=\log M(x)$. Using the following lemma, we will construct
an infinite binary sequence such that the randomness deficiency
of its initial segments grows arbitrarily slowly.
\begin{lemma}\label{deff-1}
For any set of binary strings $A$ and for any string $x$, a string $y\in A$
exists such that
$
d(xy^n)\le d(x)-\log B_{1/2}(\tilde A)+1
$
for all $n$ such that $1\le n\le l(y)$, where
$\tilde A=\cup\{\Gamma_y:y\in A\}$.
\end{lemma}
{\it Proof}. Define
$
A_1=\left\{y\in A : \exists j
(1\le j\le l(y)\&M(xy^j)>2M(x)/B_{1/2}(A))\right\}.
$
For any $y\in A_1$, denote $y^p$ be the initial fragment of $y$
of maximal length such that $M(xy^p)>2M(x)/B_{1/2}(A)$.
The set $\{y^p : y\in A_1\}$ is prefix free. Then we have
\begin{eqnarray*}
1\ge\sum\limits_{y\in A_1}\frac{M(xy^p)}{M(x)}2^{-l(y^p)}\ge
\frac{2}{B_{1/2}(\tilde A)}\sum\limits_{y\in A_1}2^{-l(y^p)}\ge
\frac{2B_{1/2}(\tilde A_1)}{B_{1/2}(\tilde A)}.
\end{eqnarray*}
From this we obtain $B_{1/2}(\tilde A_1)\le\frac{1}{2}B_{1/2}(\tilde A)$
and $B_{1/2}(\tilde A\setminus \tilde A_1)>\frac{1}{2}B_{1/2}(\tilde A)$.

For any $y\in A\setminus A_1$, we have
$
M(xy^j)\le 2M(x)/B_{1/2}(\tilde A)
$
for all $x$ such that $l(x)\le j\le(y)$.
$\bigtriangleup$

We will use the construction of Section~\ref{sec-1} to show that
that an infinite binary sequence $\omega$ exists such that
$d(\omega^n)\le\sigma(n)$ for all $n$ and the limit
(\ref{e-1a}) does not exist for the name
$\chi(\omega)\chi(T\omega)\chi(T^2\omega)\dots$ of its trajectory,
where $\chi(\omega)=i$ if $\omega\in\pi_i$, $i=0,1$. More precise, we prove that
\begin{eqnarray}
\limsup\limits_{n\to\infty}\frac{1}{n}\sum\limits_{i=0}^{n-1}\chi(T^i\omega)\ge
1/16,
\label{iineq-1}
\\
\liminf\limits_{n\to\infty}\frac{1}{n}
\sum\limits_{i=0}^{n-1}\chi(T^i\omega)\le 2r,
\label{iineq-2}
\end{eqnarray}
where $r$ is sufficiently small and the indicator function $\chi$
is defined above.

We will define by induction on steps~$s$ a sequence $\omega$
as the union of an increasing sequence of initial fragments
\begin{equation} \label{alpha-1s}
\omega(0)\subset\dots\subset\omega(k)\subset\dots
\end{equation}
We also define an auxiliary sequence of steps $s(-1)=s(0)=0<s(1)<\dots$.

Using Lemma~\ref{deff-1},
define $\omega(0)$ such that $d(\omega(0)^j)\le 2$
for all $j\le l(\omega(0))$.

Let us consider intervals of type $[a,a+2^{-n}]$
with dyadically rational endpoints, where
$a=\sum_{1\le i\le n}x_i2^{-i}$ and $x_i\in\{0,1\}$ for all $i$.
Any such interval corresponds to the binary interval
$\Gamma_x=\{\omega\in\Omega:x\subset\omega\}$ in $\Omega$,
where $x=x_1\dots  x_n$.

{\it Induction hypotheses.} Suppose that a number $k>0$, a binary sequence
$\omega(0)\subset\dots\subset\omega(k-1)$ of strings,
and a sequence of integer numbers $s(-1)=s(0)=0<s(1)<\dots<s(k-1)$
be already defined.

Suppose also, that the interval with dyadically rational endpoints
corresponding to the string
$\omega(k-1)$ is a subset of the support of the gadget $\Pi_{s(k-1)}$.
By the construction $w(\Pi_{s(k-1)})\le 2^{-h_{s(k-1)}}$. Then
$l(\omega(k-1))>h_{s(k-1)}$.

We also suppose that $d(\omega(k-1))\le\sigma(h_{s(k-2)})-5$
if $k$ is odd and $d(\omega(k-1))\le\sigma(h_{s(k-2)})$ if $k$ is even.

Consider an odd $k$. Denote $a=\omega(k-1)$ and let $I_a$ be the interval
with dyadically rational endpoints corresponding to $a$.

By the ergodic theorem the total measure of all points of $I_a$
generating $\Pi_s$-trajectories with frequency $r$ of visiting
the element $\pi_1$ tends to $2^{-l(a)}$ as $s\to\infty$.

Let $s$ be sufficiently large such that the total measure of all points
of $I_a$ generating $\Pi_s$-trajectories with frequency $\le 2r$
of visiting the element $\pi_1$ is at least $(1/2)2^{-l(a)}$.

Consider a subset of these points locating in the lower half of the
gadget $\Pi_s$. The measure of this set is at least $(1/4)2^{-l(a)}$.
By construction this set is a union of intervals $[r_1,r_2]$
from the gadget $\Pi_{s-1}$.
Easy to see that any interval $[r_1,r_2]$ of real numbers contains
a subinterval with dyadically rational endpoints of
length at least $\frac{1}{4}(r_2-r_1)$. Any such subinterval corresponds
to a binary string $b$. Let $C_a$ be a set of such strings $b$.
The Bernoulli measure of $C_a$ is at least $(1/16)2^{-l(a)}$.

Fix some such $s$ and define $s(k)=s$.

By Lemma~\ref{deff-1} an $b\in C_a$ exists such that
$d(b^j)\le d(a)+5$ for each $l(a)\le j\le l(b)$. Define $\omega(k)=b$.
By induction hypothesis,
$d(a)\le\sigma(h_{s(k-2)})-5$ and $l(a)\ge h_{s(k-1)}$.
Then $d(b^j)\le\sigma(h_{s(k-2)})<\sigma(h_{s(k-1)})
\le\sigma(l(a))\le\sigma(j)$ for all
$l(a)\le j\le l(b)$. Also, since $w(\Pi_s)\le 2^{-h_s}$, we have
$l(b)\ge h_{s(k)}$.
Therefore, the induction hypotheses and condition (\ref{iineq-2}) are valid
for the next step of induction.

Let $k$ be even. Put $b=\omega(k-1)$ and $s(k)=s(k-1)+1$. Denote $s=s(k)$.

Let us consider an arbitrary column of the gadget $\Delta_{s-1}$.
Divide all its intervals into two equal parts: upper half and lower half.
Any interval of the lower half of $\Delta_{s-1}$
generates a trajectory of length $\ge M/2$, where $M$
is the height of the gadget $\Delta_{s-1}$.
The uniform measure of the union of such subintervals is
$\frac{1}{2}\lambda(\hat\Delta_{s-1})$.
These intervals contain a set ${\cal I}_b$ of subintervals with dyadical
rational endpoints of measure at least $\frac{1}{8}\lambda(\hat\Delta_{s-1})$.

By Hoeffding inequality (\ref{strong-1g}) the measure
of all points of support of the gadget $\Delta_{s-1}$ whose
trajectories have length $\ge M/2$ and frequency of ones $\le 1/4$
is less than
$
2^{-\frac{1}{16}M}\lambda(\hat\Delta_{s-1})\le
\frac{1}{16}\lambda(\hat\Delta_{s-1})
$
(we consider sufficiently large $k$ and $M$).
Then all intervals from ${\cal I}_b$ generating trajectories
with frequency of ones more than $1/4$
have total measure at least $\frac{1}{16}\lambda(\hat\Delta_{s-1})$.

Let $\Pi_s$ is the gadget generated by the $R_s$-fold independent
cutting and stacking of the gudget $\Pi_{s-1}\cup\Delta''$.
By the construction
\begin{eqnarray}\label{vol-2s}
\gamma=\frac{\lambda(\hat\Delta'')}{\lambda(\hat\Pi_{s-1})}=
\frac{\lambda(\hat\Delta_{s-1})}{2\lambda(\hat\Pi_{s-1})}
=\frac{2^{-s+1}r}{1-2^{-s+2}}>
\nonumber
\\
>2^{-s+1}r\ge 2^{-(\sigma(h_{s-1})-\sigma(h_{s-2})+12)}.
\end{eqnarray}
Consider a set of all binary strings correspondent to intervals
from the lower half of $\Pi_{s-1}$ such that trajectories starting from these
intervals pass through an upper subcolumn of the
gadget $\Delta''$ and have frequencies of ones at least $1/4$.
Notice that any copy of the gadget $\Delta''$ has the same frequency
characteristics of trajectories.

Let $D_b$ be a set of all binary strings correspondent to these intervals.
By definition trajectory of any such interval has length at most $2M$ and
its name has at least $M/4$ ones. Hence, frequency of ones
in the name of any such trajectory is at least $\frac{1}{8}$.

Total measure of all such intervals is
at least $\frac{\gamma}{16}2^{-l(b)}$. By
Lemma~\ref{deff-1} an $c\in D_b$ exists such that
$
d(c^j)\le d(b)+1-\log\frac{\gamma}{32}\le
d(b)+(\sigma (h_{s-1})-\sigma (h_{s-2})-12)+6\le
\sigma (h_{s-1})-6<\sigma (h_{s(k-1)})-5
$
for all $j$ such that $l(b)\le j\le l(c)$. Here we have used
induction hypothesis, the inequality
$d(b)\le\sigma (h_{s(k-2)})\le\sigma(h_{s-2})$ and
the inequality (\ref{vol-2s}). Besides, by induction hypothesis
$l(b)\ge h_{s-1}$. Therefore,
$
d(c^j)<\sigma(h_{s-1})\le\sigma(l(b))\le\sigma(j)
$
for all $j$ such that $l(b)\le j\le l(c)$. Define $\omega(k)=c$.
It is easy to see that induction hypotheses are valid for this $k$.

An infinite sequence $\omega$ is defined by a sequence of its initial
fragments (\ref{alpha-1s}). We have proved that
$d(\omega^j)\le\sigma(j)$ for all $j$.

By the construction there are infinitely many initial fragments
of trajectory of the sequence $\omega$ with frequency of ones $\ge 1/16$
in their names. Also, there are infinitely many initial fragments
of this trajectory with frequency of ones $\le 2r$.
Hence, the condition~(\ref{iineq-1}) holds.
$\bigtriangleup$

The proof of Theorem~\ref{th-1b} is more complicated. Consider a sequence
of pairwise disjoint subintervals $J_i$ of unit interval $[0,1]$
of lengths $2^{-i}$, $i=1,2,\dots$ and a uniform computable sequence
$\sigma_i(n)$ of all partial recursive functions (candidates for degree of
instability).
For any $i$, we apply the construction of Section~\ref{sec-1}
to the subinterval $J_i$ and to a function $\sigma_i(n)$ in order
to define a computable ergodic measure preserving
transformation $T_i$ on $J_i$ for each $i$.
The needed transformation is defined as union of all these transformations
$T_i$. We omit details of this construction.

\subsection{Instability of universal compression schemes}

Note that an infinite sequence $\omega$ is Martion-L\"of random with respect to
a computable measure $P$ if and only if
$
Km(\omega^n)=-\log P(\omega^n)+O(1)
$
as $n\to\infty$.

Recent result of Hochman~\cite{Hoch2009} implies
an algorithmic version of the Shannon--McMillan--Breiman theorem
for Martin-L\"of random sequences:
for any computable stationary ergodic measure $P$ with entropy $H$,
$Km(\omega^n)\ge -\log P(\omega^n)-O(1)$ as $n\to\infty$ implies
\begin{eqnarray}
\lim\limits_{n\to\infty}\frac{Km(\omega^n)}{n}=
\lim\limits_{n\to\infty}\frac{-\log P(\omega^n)}{n}=H
\label{kol-comp-1}
\end{eqnarray}

The construction given in Section~\ref{sec-1} shows also an instability property
of the relation (\ref{kol-comp-1}) (this was first shown in~\cite{Vyu2003}).
\begin{theorem} \label{theorem-1}
Let $\sigma(n)$ as in Theorem~\ref{th-1} and $\epsilon$ be a sufficiently small
positive real number. A computable stationary ergodic measure $P$
with entropy $0<H\le\epsilon$ and an infinite binary sequence $\omega$
exist such that
\begin{eqnarray}
Km(\omega^n)\ge-\log P(\omega^n)-\sigma(n)
\label{def-lim-1}
\end{eqnarray}
for all $n$ and
\begin{eqnarray}
\limsup\limits_{n\to\infty}\frac{Km(\omega^n)}{n}\ge\frac{1}{4},
\label{limsup-1}
\\
\liminf\limits_{n\to\infty}\frac{Km(\omega^n)}{n}\le\epsilon.
\label{liminf-1}
\end{eqnarray}
\end{theorem}

By a prefix-free code we mean a computable sequence of one-to-one functions
$\{\phi_n\}$ from $\{0,1\}^n$ to a prefix-free set of finite sequences.
In this case a decoding method $\hat\phi_n$ also exists such that
$\hat\phi_n(\phi_n(\alpha))=\alpha$ for each $\alpha$ of length $n$.

A code $\{\phi_n\}$ is called {\it universal coding scheme} with respect to a class
of stationary ergodic sources if for any computable stationary ergodic
measure $P$ (with entropy $H$)
\begin{eqnarray}\label{asym-zv}
\lim_{n\to\infty}\frac{l(\phi_n(\omega^{n}))}{n}=H\mbox{ almost surely. }
\end{eqnarray}
Lempel--Ziv coding scheme is an example of such universal coding scheme.

We have also an instability property for any universal coding schemes.
\begin{theorem}\label{theorem-2}
Let $\sigma(n)$ and $\epsilon$ be as in Theorem~\ref{th-1}.
A computable stationary ergodic measure $P$ with entropy $0<H\le\epsilon$
exists such that for each universal code $\{\phi_n\}$ an infinite binary
sequence $\omega$ exists such that
\begin{eqnarray*}
Km(\omega^n)\ge-\log P(\omega^n)-\sigma(n)
\end{eqnarray*}
for all $n$ and
\begin{eqnarray}
\limsup\limits_{n\to\infty}\frac{l(\phi_n(\omega^{n}))}{n}\ge\frac{1}{4},
\label{limsup-1mb}
\\
\liminf\limits_{n\to\infty}\frac{l(\phi_n(\omega^{n}))}{n}\le\epsilon.
\label{liminf-1mb}
\end{eqnarray}
\end{theorem}
The proof of these theorems is based on the construction
of Section~\ref{sec-1}. For further details we refer reader
to V'yugin~\cite{Vyu2003}.



\end{document}